\documentclass[10pt,journal,epsfig]{IEEEtran}

\usepackage[dvips]{graphicx}
\usepackage{graphicx}

\usepackage{amssymb}
\usepackage{cite}
\usepackage{amsmath}
\usepackage{algorithm}
\usepackage{algorithmic}
\usepackage{multirow}
\usepackage[table]{xcolor}
\usepackage{subfigure}
\usepackage{makecell}
\usepackage{diagbox}
\usepackage{array}
\usepackage{threeparttable}
\usepackage{graphicx}
\usepackage{caption,epstopdf}
\usepackage{color}
\providecommand{\keywords}[1]{\textbf{\textit{Index terms---}} #1}

\begin{document}

\title{CSI-Free Position Optimization for Movable Antenna Communication Systems: A Black-Box
    Optimization Approach}

\author{Xianlong Zeng, Jun Fang, Bin Wang, Boyu Ning, and Hongbin Li,~\IEEEmembership{Fellow,~IEEE}
\thanks{Xianlong Zeng, Jun Fang, Bin Wang and Boyu Ning are with the National Key Laboratory
of Wireless Communications, University of Electronic Science and
Technology of China, Chengdu 611731, China, Emails:
JunFang@uestc.edu.cn; boydning@outlook.com}
\thanks{Hongbin Li is with the Department of Electrical and Computer Engineering,
Stevens Institute of Technology, Hoboken, NJ 07030, USA, Email:
Hongbin.Li@stevens.edu}
 } \maketitle

\begin{abstract}
Movable antenna (MA) is a new technology which leverages local
movement of antennas to improve channel qualities and enhance the
communication performance. Nevertheless, to fully realize the
potential of MA systems, complete channel state information (CSI)
between the transmitter-MA and the receiver-MA is required, which
involves estimating a large number of channel parameters and
incurs an excessive amount of training overhead. To address this
challenge, in this paper, we propose a CSI-free MA position
optimization method. The basic idea is to treat position
optimization as a black-box optimization problem and calculate the
gradient of the unknown objective function using zeroth-order (ZO)
gradient approximation techniques. Simulation results show that
the proposed ZO-based method, through adaptively adjusting the
position of the MA, can achieve a favorable signal-to-noise-ratio
(SNR) using a smaller number of position measurements than the
CSI-based approach. Such a merit makes the proposed algorithm more
adaptable to fast-changing propagation channels.
\end{abstract}


\section{Introduction}
{Movable antenna (MA) \cite{zhu23} and fluid antenna (FA) \cite{wongZhang20} are emerging technologies that have drawn widespread attention for its ability of allowing antennas movable over a specified region to obtain a better channel quality.
As compared to the traditional fixed-position antennas, MA/FA is able to fully exploit the spatial DoFs with much fewer antennas or even a single antenna \cite{zhuZhang23}.}
 
Over the past few years, many efforts have been made to investigate the potential of the MA systems, e.g. \cite{zhuMa23,meizhang2024,artner23,chenSchober2023,xiao23,ningYang24}. 
{Among them, the work \cite{zhuMa23,meizhang2024,artner23} proposed to optimize the positions of MA to improve the spatial multiplexing performance and the received signal power. In \cite{chenSchober2023}, an MA-enhanced MIMO system was introduced to jointly optimize antenna positions and the transmit covariance matrix to maximize the achievable rate based on statistical CSI.} The MA-aided interference suppression problem was investigated in \cite{xiao23}, which showed that MA-assisted
multiuser systems can not only increase the receive signal power, but also achieve effective interference mitigation.
In addition, the work \cite{ningYang24} discussed general architectures 
and implementation methods for realizing MA in existing communication systems. 
{In addition to the above works, the potential application of MA/FA to mobile edge computing (MEC) and other fields was also studied, e.g. \cite{zuoJin24}.}

Despite the advantages of MA systems over fixed-position antenna
(FPA) systems, the operation of the MA system relies on the
complete knowledge of the channel response between the transmitter
(Tx) and the receiver (Rx) over the entire movable regions
\cite{xiaoCao24}. If channel measurement is taken for every point
over the movable region in order to find a best MA position, this
would undoubtedly involve an excessive amount of time and training
overhead. To address this issue, a compressed sensing-based method
was proposed in \cite{maZhu23} by exploiting the multi-path field
response channel structure. Based on multiple measurements taken
at designated positions of the Tx-MA and Rx-MA, the channel
parameters associated with multi-path components can be estimated
via a compressed sensing approach and the channel response over
the entire movable regions can be reconstructed. Such a compressed
sensing-based approach, however, faces difficulties under
rich-scattering channels with a large number of multi-path
components.

In this paper, we propose a CSI-free position optimization
approach for MA-assisted communication systems. The idea is to
treat position optimization as a black-box optimization problem,
and estimate the gradient of the objective function via
zeroth-order (ZO) gradient approximation methods. Specifically,
based on the received signals collected from previous positions,
the receiver calculates a new position and the MA is then moved to
this new position to take measurements for next position
refinement. Such a procedure is repeated until a convergence is
reached. Simulation results show that the proposed method is more
sample-efficient than the CSI-based method in optimizing the MA's
position. Moreover, another advantage of our proposed approach it
only utilizes the magnitude of the received signals for position
optimization, which enhances the algorithm's robustness against
phase noise/errors inevitably present in the receiver system.

\begin{figure}[t!]
    \setlength{\abovedisplayskip}{1pt}
    \setlength{\belowdisplayskip}{1pt}
    \centering
    \includegraphics[width=4.4cm]{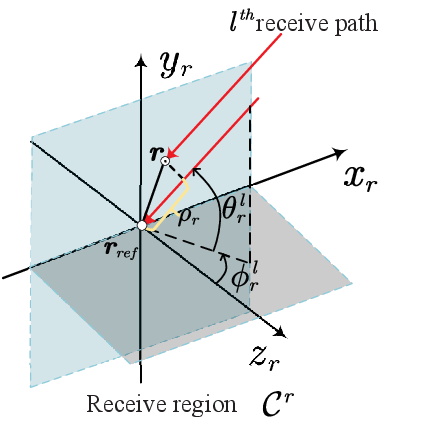}
    \caption{Schematic of Rx-MA's coordinates and geometric relationships}
    \label{fig:system}
\end{figure}

\section{System Model}
We consider a movable antenna (MA)-aided point-to-point narrowband
communication system, where a single MA is employed at the
receiver, and a single fixed-position antenna is employed at the
transmitter. Such a simple system model helps facilitate the
exposition of the idea of our work. Our proposed scheme can also
be extended to the joint Tx-Rx MA system at the cost of involving
additional feedback from the Rx to the Tx.

As illustrated in Fig. \ref{fig:system}, we establish a local
Cartesian coordinate system to describe the position of the
receive antenna. The position of the RX's antenna is denoted as
$\boldsymbol{r}=[x^r,y^r] \in \mathcal{C}^r$, where
$\mathcal{C}^r$ is a two-dimensional square region of size
$A\times A$ in which the antenna can be flexibly positioned.

Denoting the channel from the Tx to the Rx as $h(\boldsymbol{r})$,
the received signal at the Rx can be expressed as
\begin{align}
    y(\boldsymbol{r})=\sqrt{P}h(\boldsymbol{r})s+n, \label{recieve}
\end{align}
where $s$ denotes the transmitted signal, $P$ is the transmit
power and $n$ denotes the additive white Gaussian noise with zero
mean and variance $\sigma^2$.

Note that the channel response $h$ between the Tx and the Rx is a
superposition of multi-path components (MPCs). Specifically, it
can be modeled by the geometric channel model as
\begin{align}
    h&=\sum_{l=1}^{L}b_l=\sum_{l=1}^{L}\alpha_l e^{j\delta_l}, \label{channel}
\end{align}
where $b_l\triangleq\alpha_l e^{j\delta_l}$ represents the complex
coefficient associated with the $l$th path, $\alpha_l$ and
$\delta_l$ respectively represent the magnitude and the phase of
$b_l$. The above channel response model, however, does not take
the position of the antenna into account. To model the effect of
the MA's position, we first consider (\ref{channel}) as the
channel response from the Tx to the receive reference position
$\boldsymbol{r}_{ref}=[0,0]^T$, and express $h$ as
\begin{align}
    h(\boldsymbol{r}_{ref})=\boldsymbol{1}_{L}^H\boldsymbol{\varGamma },
\end{align}
where $\boldsymbol{\varGamma }\triangleq [b_1\phantom{0}
b_2\phantom{0}\cdots\phantom{0} b_L]^T =[\alpha_1
e^{j\delta_1}\phantom{0}\cdots\phantom{0}\alpha_L e^{j\delta_L}]^T
\in\mathbb{C} ^{L}$ denotes the path response vector and
$\boldsymbol{1}_{L}\in \mathbb{R}^{L}$ denotes a vector with all
entries equal to one.

Here, we assume that the far-field condition is satisfied between
the Tx and the Rx, where the size of $\mathcal{C}^r$ is much
smaller than the propagation distance. Hence, the angles of
arrival (AoAs) and the complex coefficients of MPCs are
approximately the same over the region $\mathcal{C}^r$.
Nevertheless, as the antenna moves from the reference point
$[0,0]$ to position $[x_r,y_r]$, the signal propagation time of
each path changes, which in turn results in a phase variation.
Specifically, as illustrated in Fig. \ref{fig:system}, denote
$\theta^{l}_r$ and $\phi^{l}_r$ as the $l$-th path's elevation AoA
and azimuth AoA, respectively. When the { Rx-MA} moves to position
$\boldsymbol{r}=[x_r,y_r]$ from $\boldsymbol{r}_{ref}=[0,0]$, the
signal propagation distance for the $l$-th path is changed by
\begin{align}
    \rho^{l}_{r}(x_r,y_r)
    =x_r \cos\theta^{l}_r\sin \phi^{l}_r+y_r \sin \theta^{l}_r,
\end{align}
As a result, the $l$-th path incurs a phase variation of
$\frac{2\pi}{\lambda} \rho^{l}_{r}(x_r,y_r)$ at position
$\boldsymbol{r}=[x_r,y_r]$ with respect to position
$\boldsymbol{r}_{ref}=[0,0]$, where $\lambda$ denotes the wavelength
of the signal. Therefore, the channel response between the Tx and
the { Rx-MA} located at $\boldsymbol{r}=[x_r,y_r]$ can be expressed
as
\begin{align}
h(\boldsymbol{r})=\sum_{l=1}^{L}b_le^{-j\frac{2\pi}{\lambda}
\rho^{l}_{r}(x_r,y_r)}
=\boldsymbol{f}(\boldsymbol{r})^H\boldsymbol{\varGamma }
\label{matrix},
\end{align}
where $\boldsymbol{f}(\boldsymbol{r})\in \mathbb{C}^{L}$ denotes the
receiver-side field response vector accounting for the phase
variations, which is given by
\begin{align}
    \boldsymbol{f}(\boldsymbol{r})&=\left[e^{\frac{2\pi}{\lambda}\rho^{1}_{r}(x_r,y_r)}\phantom{0}
    \cdots\phantom{0} e^{\frac{2\pi}{\lambda}\rho^{L}_{r}(x_r,y_r)}
    \right]^T.\nonumber
\end{align}
We see that the channel response between the Tx and the Rx is
dependent on the position of the Rx-MA. As a result, through
changing/configuring the position of the MA, a better performance,
e.g. a higher signal-to-noise ratio, can be obtained.

\section{Problem Formulation}
To evaluate the performance of MA-aided communication systems, we
adopt the signal-to-noise ratio (SNR) as a metric. According to
(\ref{recieve}), the SNR for the received signal is given by
\begin{align}
    \gamma(\boldsymbol{r})=\frac{\left| h(\boldsymbol{r})\right|^2
        P}{\sigma^2}.\label{eqn4}
\end{align}
Here the SNR is depended on the position of the { Rx-MA}.

Our objective is to maximize the receive SNR via optimizing the
Rx-MA position. Such a problem can be formulated as
\begin{align}
    (\mathrm{P1}) \quad \max_{\boldsymbol{r}} \quad &
    \gamma(\boldsymbol{r})=\frac{\left| h(\boldsymbol{r})\right|^2 P}{\sigma^2} \nonumber\\
    \text{s.t.}\quad  &\boldsymbol{r}\in\mathcal{C}_r. \label{eqn1}
\end{align}
Since the transmit power $P$ and the noise power $\sigma^2$ are
constants, problem (P1) is equivalent to maximizing $\left|
h(\boldsymbol{r})\right|^2$, which is further expressed as
\begin{align}
    \begin{aligned}
        \left| h(\boldsymbol{r})\right|^2&=\left|\sum_{l=1}^{L}b_l e^{-j\rho_r^l\left(x_r,y_r\right)}\right|^2 \\
        &=\sum_{m=1}^{L}\sum_{n=1}^{L} \left|\alpha_m\right| \left| \alpha_n\right| \cos \left(\frac{2\pi}{\lambda}
        \varrho_{mn} + c_{mn}\right).\label{power}
    \end{aligned}
\end{align}
where $\varrho_{mn}(\boldsymbol{r})\triangleq
\rho_r^m\left(x_r,y_r\right)-\rho_r^n\left(x_r,y_r\right)$ and
$c_{mn}\triangleq\delta_n-\delta_m$.

From (\ref{power}), we see that the optimization of the MA's
position requires the knowledge of the channel parameters
$\{b_l,\,\theta^{l}_r,\, \phi^{l}_r\}_{l=1}^L$. To obtain these
channel parameters, we need to measure the channel
$h(\boldsymbol{r})$ at different positions, say,
$\{\boldsymbol{r}_m\}_{m=1}^M$, and collect the measured channel
samples $\{h(\boldsymbol{r}_m)\}_{m=1}^M$. Based on the measured
channel samples $\{h(\boldsymbol{r}_m)\}_{m=1}^M$, the channel
parameters $\{b_l,\,\theta^{l}_r,\, \phi^{l}_r\}_{l=1}^L$ can be
estimated via compressed sensing or other signal processing
techniques. Such a channel estimation-based approach, however, may
require an excessively large number of channel measurements for
rich scattering scenarios consisting of a heavy number of MPCs.
To overcome the drawback of existing approaches, in this paper we
propose a CSI-free approach which does not need the knowledge of
the channel parameters for position optimization.

\section{Proposed Method}
Note that the objective function of (\ref{eqn1}) has unknown
channel parameters, making it challenging or even impossible to
compute explicit expressions of its gradient. Such optimization
problems involving unknown objective functions are known as
block-box optimization problems, and have been investigated by
many prior works. Although there are different approaches (such as
Bayesian optimization\cite{shahriariBobak15}, derivative-free
trust region methods\cite{larsonJeffrey19}, and genetic
algorithms\cite{berahasCao22}) to address the block-box
optimization problem. Among them, the zeroth-order (ZO)
optimization method \cite{chenLiu19,liuChen20} has
gained an increasing attention due to its unique advantages:
first, ZO methods are easy to implement as they are based on
commonly used gradient-based algorithms; second, ZO methods can
achieve comparable convergence rates to first-order algorithms.

\subsection{ZO Optimization Method}
The basic idea of ZO optimization is to approximate the full
gradients or stochastic gradients through function value-based
gradient estimates. Specifically, for an unknown function $f$, the
ZO gradient is estimated as the central difference of two function
values at a random unit direction \cite{liuChen20}:
\begin{align}
    \hat{\nabla}
    f(\boldsymbol{x})=(d/2\mu)[f(\boldsymbol{x}+\mu\boldsymbol{u})-f(\boldsymbol{x}-\mu\boldsymbol{u})]
    \boldsymbol{u},\label{eqn2}
\end{align}
where $\boldsymbol{u}$ is a random vector drawn from the sphere of
a unit ball, $\mu$ is a small step size and $d$ denotes the dimension of the variable $\boldsymbol{x}$. In some cases, $\boldsymbol{u}$ can be
also randomly chosen as a standard unit vector $\boldsymbol{e}_i$
with 1 at its $i$th coordinate and zeros elsewhere.

Most ZO optimization methods mimic their first-order counterparts
and involve three steps, namely, gradient estimation, descent
direction computation, and point updating. The gradient estimation
can be performed using (\ref{eqn2}). For different ZO methods,
their major difference lies in the strategies used to form the
descent direction. Note that the estimated gradient (\ref{eqn2})
is stochastic in nature and may suffer from a large estimation
variance, which causes poor convergence performance. To address
this issue, different descent direction update schemes were
proposed. Among them, the ZO-AdaMM \cite{chenLiu19} has been
proven to be an effective method that is robust against gradient
estimation errors and achieves a superior convergence speed. Due
to its simplicity and superior performance, here we adopt ZO-AdaMM
to solve our position optimization problem.

For clarity, we provide a summary of ZO-AdaMM in Algorithm
\ref{zoadamm1}.
\begin{algorithm}[htb]
    \caption{ZO-AdaMM }
    \label{zoadamm1}
    \begin{algorithmic}[1]
        \REQUIRE  step size $\alpha$, hyper-parameters
        $\beta_{1}$, $\beta_2\in \left(0,1\right] $, and set $\boldsymbol{m}_0$, $\boldsymbol{v}_0$ ~~\\
        \STATE { Initialize the MA position $\boldsymbol{r}_{0}$.}
        \WHILE {not converge}
        \STATE {$t \gets t +1 $ }
        \STATE {Estimate $\tilde{\boldsymbol{g}}_t= \hat{\nabla}f(\boldsymbol{x})$  according to (\ref{eqn2})}

        \STATE {let     $\boldsymbol{m}_{t}=\beta_1 \boldsymbol{m}_{t-1}
            +(1-\beta_1)\tilde{\boldsymbol{g}}_t$} \STATE {
            $\boldsymbol{v}_{t}=\beta_2 \boldsymbol{v}_{t-1}
            +(1-\beta_2)\tilde{\boldsymbol{g}}_t\circ\tilde{\boldsymbol{g}}_t$\\}
        \STATE {    $\hat{\boldsymbol{m}}_{t}=\frac{\boldsymbol{m}_{t}}{1-\beta_1^t}$}
        \STATE {
            $\hat{\boldsymbol{v}}_{t}=\frac{\boldsymbol{v}_{t}}{1-\beta_2^t}$,
            and
            $\hat{\boldsymbol{V}}_{t}=\text{diag}(\hat{\boldsymbol{v}}_{t})$ }
        \STATE {Update $\boldsymbol{r}_{t+1}\gets \boldsymbol{r}_{t}+
            \alpha\hat{\boldsymbol{V}}_{t}^{-\frac{1}{2}}{\hat{\boldsymbol{m}}_{t}}$
        }
        \ENDWHILE
        \ENSURE $\boldsymbol{r}^{*}$\\
    \end{algorithmic}
\end{algorithm}

In summary, the ZO-AdaMM is a stochastic gradient descent method
that consists of three steps, namely, estimation of the ZO
gradient, descent direction and learning rate calculation, and
variable (i.e. position) update.

First, for each position $\boldsymbol{r}$, the ZO-AdaMM employs
(\ref{eqn2}) to compute an estimate of the ZO gradient:
\begin{align}
    \begin{aligned}
        \tilde{\boldsymbol{g}} = \hat{\nabla}       f(\boldsymbol{x})
        \approx \frac {d(\gamma(\boldsymbol{r}+\mu \boldsymbol{u}) -
            \gamma(\boldsymbol{r}-\mu \boldsymbol{u}))}{2\mu}\boldsymbol{u}\\
        \overset{(a)}{=}\frac {d\left(\left|h(\boldsymbol{r}+\mu \boldsymbol{u})\right|^2 -
            \left|h(\boldsymbol{r}-\mu \boldsymbol{u})\right|^2\right)}{2\mu
           }\boldsymbol{u} \\
\approx \frac {d\left(\left|y(\boldsymbol{r}+\mu
\boldsymbol{u})\right|^2 -
            \left|y(\boldsymbol{r}-\mu \boldsymbol{u})\right|^2\right)}{2\mu
           }\boldsymbol{u},
    \end{aligned}\label{ZO-AP}
\end{align}
where $\tilde{\boldsymbol{g}}$ denotes the estimated gradient
vector, and in $(a)$, we ignore the scaling term $P/\sigma^2$ as
it is independent of the optimization variable. In our algorithm,
the vector $\boldsymbol{u}$ is randomly chosen as a standard unit
vector for each iteration, Note that, since the channel
$h(\boldsymbol{r})$ is not directly available, we use its noisy
version $y(\boldsymbol{r})$ (cf. (\ref{recieve})) to calculate the
gradient. On the other hand, the gradient estimate
$\tilde{\boldsymbol{g}}$ becomes more accurate when a smaller
$\mu$ is adopted. However, if $\mu$ is set too small, the function
difference could be dominated by the noise and thus yields a poor
gradient estimate. Thus, a proper choice of the parameter $\mu$ is
important for the convergence of ZO-AdaMM.

In the second step of ZO-AdaMM, we need to calculate the descent
direction and the adaptive learning rate. The descent direction is
given by an exponential moving average of past gradients.
Specifically, at each iteration, the descent direction can be
updated as
\begin{align}
\boldsymbol{m}_{t}=\beta_1 \boldsymbol{m}_{t-1}
    +(1-\beta_1)\tilde{\boldsymbol{g}}_t,
\end{align}
where the hyperparameter $\beta_1\in \left[0,\,\,1 \right)$
controls the exponential decay rate of the moving averages. In
particular, the descent direction is not only determined by the
gradient at the current iteration, but also depends on past
gradients. The second moment vector is adaptively calculated as
\begin{align} 
\boldsymbol{v}_{t}=\beta_2
    \boldsymbol{v}_{t-1}+(1-\beta_2)\tilde{\boldsymbol{g}}_t\circ\tilde{\boldsymbol{g}}_t,
\end{align}
where the hyperparameter $\beta_2\in\left[0,\,\,1 \right)$
controls the exponential decay rate of the moving averages and
$\circ$ denotes the Hadamard product.

Typically, $\boldsymbol{m}_{0}$ and $\boldsymbol{v}_{0}$ are
initialized as zeros. This results in moving averages
$\boldsymbol{m}_{t}$ and $\boldsymbol{v}_{t}$ that are biased
towards zero, especially during the initial few iterations. This
bias can be easily counteracted, by adopting bias-corrected
estimates $\hat{\boldsymbol{m}}_{t},\hat{\boldsymbol{v}}_{t}$,
i.e.,
\begin{align}
\hat{\boldsymbol{m}}_{t}=\frac{\boldsymbol{m}_{t}}{1-\beta_1^t} \qquad
\hat{\boldsymbol{v}}_{t}=\frac{\boldsymbol{v}_{t}}{1-\beta_2^t}
\end{align}
At the beginning, the values of $\boldsymbol{m}_{t}$ and
$\boldsymbol{v}_{t}$ are small because they are weighted averages
of estimated gradients, and the initial values are set as $0$.
Nevertheless, we can amplify the $\boldsymbol{m}_{t}$ and
$\boldsymbol{v}_{t}$ by the bias correction factor
$\{1-\beta_i^t,i=1,2\}$ to make them closer to the actual moving
averages. In particular, as the number of iterations increases,
the value of $\beta_i^t$ converges to 0 and the bias correction
factor will finally approach to 1, so that the effect of the bias
correction will gradually diminish.

After the descent direction and the second moment vector are
calculated, the MA position can be updated as
\begin{align}
\boldsymbol{r}_{t+1}\gets \boldsymbol{r}_{t}+
    \alpha\hat{\boldsymbol{V}}_{t}^{-\frac{1}{2}}{\hat{\boldsymbol{m}}_{t}},
\end{align}
where $\boldsymbol{V}_{t}\triangleq
\text{diag}\{\boldsymbol{v}_{t}\}$ and $\alpha$ denotes the step
size. $\alpha\hat{\boldsymbol{V}}_{t}^{-\frac{1}{2}}$ can be
interpreted as the adaptive learning rate which can dynamically
adjust the step size. Particulary, when the slope is steeper, the
step size will become smaller to avoid missing the optimal point,
which ensures the robustness of the algorithm. From another
perspective, the rationale behind the above equation is to
normalize the estimated descent direction to reduce the noise
effect. We can see that ZO-AdaMM normalizes the descent direction
$\hat{m}_{t}$ by $\sqrt{\hat{v}_{t}}$. In particular,
$\hat{m}_{t}/\sqrt{\hat{v}_{t}}$ is invariant to the scale of the
gradients. Scaling the gradient with factor $c$ will scale
$\hat{m}_{t}$ with a factor $c$ and $\hat{v}_{t}$ with a factor
$c^2$, which cancel out this scaling factor as we have
${c\hat{m}_{t}}/{\sqrt{c^2\hat{v}_{t}}}={\hat{m}_{t}}/{\sqrt{\hat{v}_{t}}}$.

\subsection{Other Implementing Issues}
A good initialization point can help the ZO-AdaMM algorithm
converge to the desired minimum more quickly. To obtain a good
initialization point, we can have the Rx-MA move to different
positions to measure the channel magnitude and select the position
with the largest magnitude.

Another issue is that the algorithm may produce a position that is
out of the feasible region. In this case, we project its position
component to the corresponding edge value, i.e.,
\begin{equation}
    [\boldsymbol{r}]_i=\left\{
    \begin{aligned}
        -\frac{A}{2} \quad  &\text{if $\left[\boldsymbol{r}\right]_i <-\frac{A}{2}$},\\
        \frac{A}{2}\quad     & \text{if $\left[\boldsymbol{r}\right]_i >\frac{A}{2}$}, \\
        \left[\boldsymbol{r}\right]_i\quad  & \text{otherwise}. \\
    \end{aligned}
    \right.
\end{equation}

{Next, we analyze the computational complexity of the proposed ZO-AdaMM algorithm. The dominant computational cost comes from calculating the ZO gradient and updating the learning rate, both of which has a complexity of $\mathcal{O}(d)$. Therefore the overall computational complexity of the proposed method is in the order of $\mathcal{O}(dT)$, where $T$ denotes the number of iterations.} 

\subsection{Remarks}
Although in this paper we consider the channel model in a form of (\ref{matrix}), our proposed method is quite general and can be applied to other
forms of ${h}(\boldsymbol{r})$. Since our proposed method is essentially a black-box
optimization approach. It works without specifying any particular expression of ${h}(\boldsymbol{r})$ and 
without assuming the knowledge of the expression of
$\boldsymbol{h}(r)$.

\section{Simulation Result}
We now presents simulation results to show the efficiency of the
proposed CSI-free ZO method for position optimization. During the
training stage, the Tx equipped with an FPA transmits a constant
signal $s=1$ to the receiver. Based on the received signals
collected from previous positions
$\{y(\boldsymbol{r}_t\pm\mu\boldsymbol{u}_t)\}_{t=0}^{t'}$, the
receiver calculates a new position, $\boldsymbol{r}_{t'+1}$,
according to Algorithm 1, and then collects measurements at
positions $\boldsymbol{r}_{t'+1}\pm\mu\boldsymbol{u}_{t'+1}$ for
subsequent position update. This procedure is repeated until a
convergence is reached. Note that here $\boldsymbol{u}_t$
represents the standard unit vector randomly chosen at the $t$th
iteration.

In our experiments, the channel is generated according to
(\ref{matrix}), where the path response coefficients $\{b_l\}$ are
independent and identically distributed (i.i.d.) circularly
symmetric complex Gaussian random variables, i.e., $b_l\sim
\mathcal{CN}(0,1/L)$. The elevation AoA and the azimuth AoA
associated with each path are uniformly chosen over the interval
$[-\pi/2,\pi/2]$. 
The region in which the MA can be flexibly adjusted is
set to a square area with a size of $4\lambda\times 4\lambda$,
i.e., $\mathcal{C}_r=[-2\lambda,2\lambda]\times
[-2\lambda,2\lambda]$. {The carrier frequency is set to 5GHz.}
 For our proposed method, the
hyper-parameters $\beta_1$ and $\beta_2$ are set respectively as
$\beta_1=0.9$ and $\beta_2=0.99$. The transmit signal-to-noise ratio (SNR) is set to 
$P/\sigma^2=30$dB. The performance of the proposed method is
evaluated by the receive SNR (\ref{eqn4}) corresponding to the
MA's position obtained by the proposed method.

\begin{figure}[htb]
   \setlength{\abovedisplayskip}{1pt}
   \setlength{\belowdisplayskip}{1pt}
   \centering
   \includegraphics[width=5.00cm]{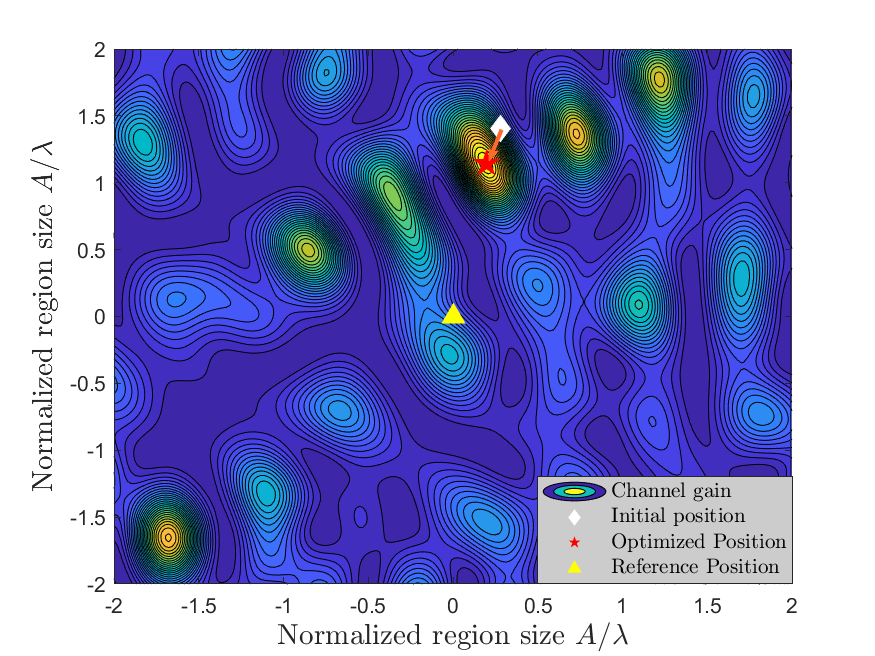}
   \caption{Variation of the receive SNR over the movable region.}
   \label{fig:1}
\end{figure}

In Fig. \ref{fig:1}, we depict the variation of the receive SNR
over the entire movable region, where we set the number of paths
$L=30$. It can be seen that, due to the small-scale fading in the
spatial domain, the channel quality varies drastically across the
movable region. In the figure, the reference point $[0,0]$ of the
MA has a receive SNR of $11.5$dB. After some iterations, the
proposed algorithm finally converges to a position
$[0.36\lambda,1.01\lambda]$, which yields a receive SNR of
$18.4$dB, and achieves an SNR increase of about $6.9$dB as
compared to the reference point.

\begin{figure}[htb]
   \begin{minipage}[b]{0.48\linewidth}
       \centering
       \centerline{\includegraphics[width=4.605cm]{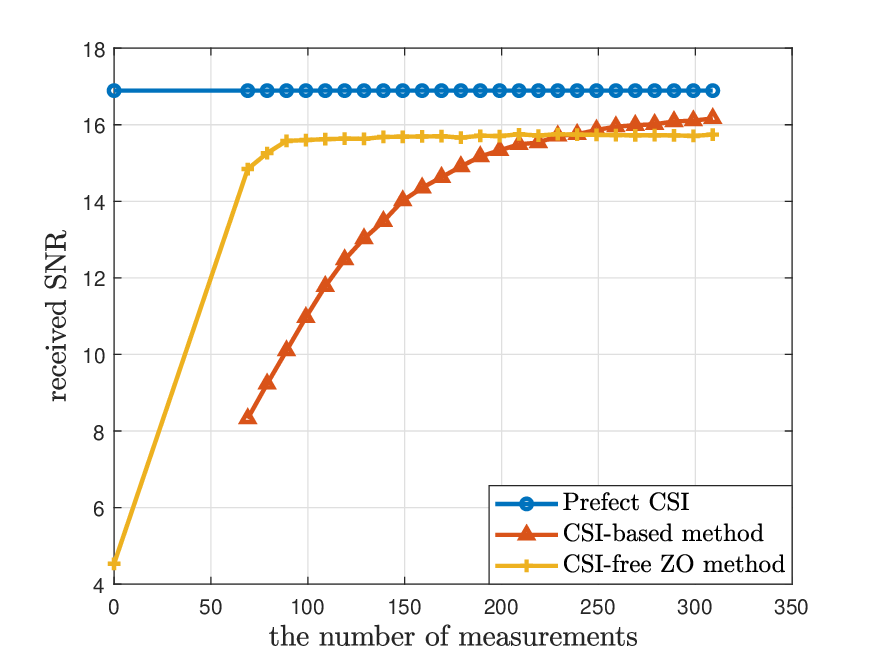}}
       \centerline{(a) }
   \end{minipage}
   \begin{minipage}[b]{0.48\linewidth}
       \centering
       \centerline{\includegraphics[width=4.605cm]{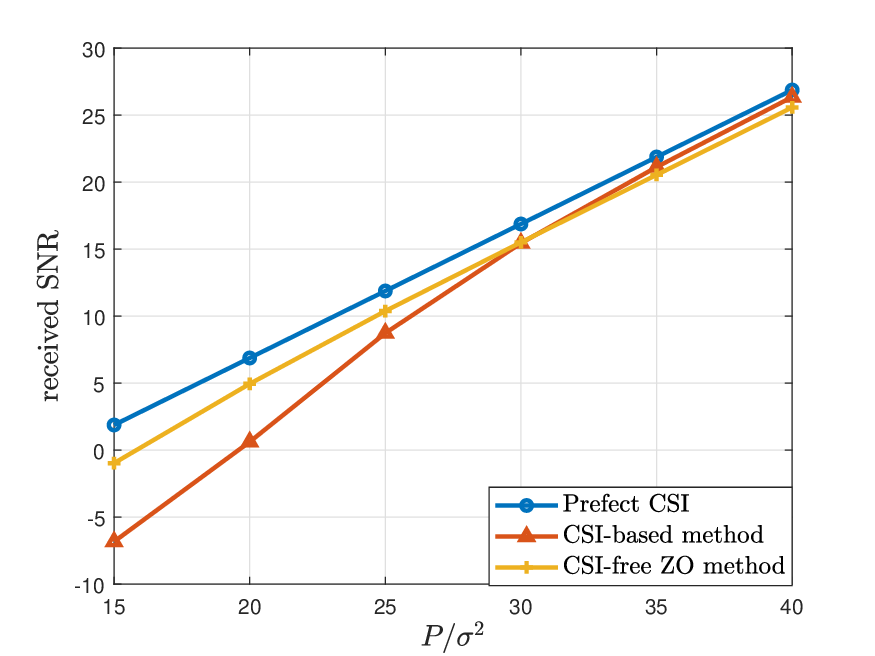}}

       \centerline{(b)}
   \end{minipage}
          \vspace{0.1cm}
   \centering{
\caption{(a) Receive SNRs achieved by respective methods; (b)
received SNRs versus the noise variance.}} \label{fig:res}
\end{figure}

Note that our proposed method requires a total number of $T'=N+2T$
channel measurements, in which the first $N$ measurements are used
to determine a good initialization point and the rest $2T$
measurements are used to calculate the ZO gradients during the
iterative process, with $2$ samples needed at each iteration for
computing the current ZO gradient. To show the sample
efficiency of the proposed algorithm, we compare our CSI-free
method with the CSI-based method \cite{maZhu23}. For the CSI-based
method, the MA needs to collect multiple channel measurements at
designated or randomly selected positions and then use a
compressed sensing-based method to estimate the channel
parameters. After the channel parameters are estimated, the
optimum position that achieves the maximum receive SNR can be
determined via a two-dimensional search.

Fig. 3(a) illustrates the receive SNRs of respective methods versus
the total number of channel measurements, where the number of
paths is set to $L=30$. We see that our proposed algorithm
presents a clear performance advantage over the CSI-based method
in terms of sampling efficiency: it requires only about $T'=69$
channel samples to find a good point that achieves a receive SNR
close to the maximum achievable SNR, whereas the CSI-based method
takes about $T'=209$ channel samples to attain a similar
performance. When the number of samples is limited, say $T'=100$,
our proposed algorithm is able to achieve a significant SNR
performance advantage over the CSI-based method.

Next, we show the performance of respective methods as a function
of the noise variance. Note that both our proposed method and the
CSI-based method rely on the received channel measurements
$y(\boldsymbol{r})$ to find a good position for the RX-MA.
Therefore it is interesting to examine the behavior of respective
algorithms when the measurements are corrupted by different
amounts of noise levels. Fig. 3 (b) depicts the receive SNRs
achieved by respective methods versus the noise variance, where
the total number of channel measurements and the number of paths
are set to $T'=209$ and $L=30$, respectively. We see that our
proposed method exhibits a better performance than the CSI-based
method when the noise level added to the measurements is high.
This improved robustness against noise comes from the fact that
our proposed method does not need to explicitly acquire the
channel parameters, instead, it simply uses the channel
measurements to calculate an estimate of the gradient of the
objective function. As the ZO-AdaMM method is robust against
estimation errors of the gradients, it explains why our proposed
method presents a performance improvement over the CSI-based
method in a low SNR regime.

\section{Conclusions}
In this paper, we proposed a CSI-free position optimization
approach for MA-assisted communication systems. The proposed
method adaptively adjusts the position of the { Rx-MA} based simply
on channel measurements, without the need of acquiring the channel
response between the Tx and the Rx over the entire movable
regions. Simulation results show that the proposed method presents
a significant performance advantage over the CSI-based method when
the number of channel measurements is limited.


\end{document}